# Influence of outer-layer finite-size effects on the rupture kinetics of a thin polymer film embedded in an immiscible matrix

M. S. Chebil[1], J. D. McGraw[2,3] [*], T. Salez[4,5], C. Sollogoub[1], G. Miquelard-Garnier[1,*]

joshua.mcgraw@espci.fr, guillaume.miquelardgarnier@lecnam.net

[1]: Laboratoire PIMM, UMR 8006, ENSAM, CNRS, CNAM, HESAM, 151 boulevard de l'Hôpital, 75013 Paris, France

[2]: Département de Physique, Ecole Normale Supérieure/PSL Research University, CNRS, 24 Rue Lhomond, 75005 Paris, France

[3]: Laboratoire de Microfluidique, MEMs et Nanostructures, UMR CNRS Gulliver 7083, ESPCI Paris/PSL Research University, 75005 Paris, France

[4]: Univ. Bordeaux, CNRS, LOMA, UMR 5798, F-33405 Talence, France

[5]: Global Station for Soft Matter, Global Institution for Collaborative Research and Education, Hokkaido University, Sapporo, Hokkaido 060-0808, Japan.

**Table of Contents entry**

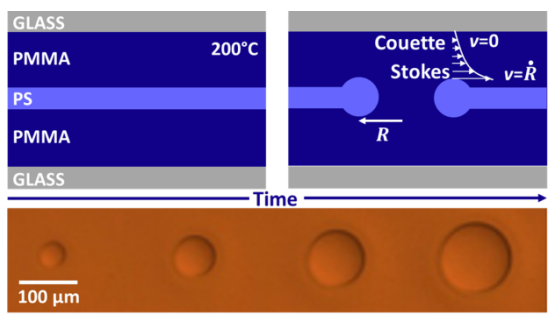

*Balancing capillary driving power and viscous dissipation with a no-slip boundary condition captures the hole growth dynamics in trilayer dewetting.*

**Abstract**

In capillary-driven fluid dynamics, simple departures from equilibrium offer the chance to quantitatively model the resulting relaxations. These dynamics in turn provide insight on both practical and fundamental aspects of thin-film hydrodynamics. In this work, we describe a model trilayer dewetting experiment elucidating the effect of solid, no-slip confining boundaries on the bursting of a liquid film in a viscous environment. This experiment was inspired by an industrial polymer processing technique, multilayer coextrusion, in which thousands of alternating layers are stacked atop one another. When pushed to the nanoscale limit, the individual layers are found to break up on time scales shorter than the processing time. To gain insight on this dynamic problem, we here directly observe the growth rate of holes in the middle layer of the trilayer films described above, wherein the distance between the inner film and solid boundary can be orders of magnitude larger than its thickness. In otherwise identical experimental conditions, thinner films break up faster than thicker ones. This observation is found to agree with a scaling model that balances capillary driving power and viscous dissipation with, crucially, a no-slip boundary condition at the solid substrate/viscous environment boundary. In particular, even for the thinnest middle-layers, no finite-size effect is needed to explain the data. The dynamics of hole growth is captured by a single master curve over four orders of magnitude in the dimensionless hole radius and time, and is found to agree well with predictions including analytic expressions for the dissipation.

**Introduction**

The study of dynamics of thin liquid films involves both fundamental and applied research.[1-3] Furthermore, the evolution of such films, by virtue of the fact that they exhibit a relatively high surface-to-volume ratio, are particularly sensitive to interfacial phenomena. These films thus present an ideal testing ground for the associated interfacial physics[4-11], and such studies have also proven useful to quantify processes such as coating,[12] or to develop alternative processing strategies to nanolithography.[13] The stability and dynamics of thin polymer films in particular has been intensively studied over the past 25 years, both theoretically[14-19] and experimentally.[20-27] The ongoing attention is also due in part to the fact that thin liquid films and even thin solid coatings are prevalent in many commercial, industrial and biological settings.

When a thin (typically below ~1 µm) polymer film is deposited on top of a substrate that does not form a completely wetting layer in equilibrium,[28] the film may dewet when heated above its glass transition temperature, $T_g$. In this process, the underlying substrate is progressively exposed to the surrounding atmosphere by the growth of a hole, reducing the contact between the liquid and the substrate. Dewetting thus reduces the total interfacial energy. Analogously, a thin film may instead lie on an immiscible, liquid polymer substrate and the same hole growth process can similarly reduce the interfacial energy. Several regimes have been predicted[17, 29, 30] for the dewetting dynamics in such systems and have been observed experimentally.[16, 21, 22, 31] Specifically, it was shown that for a liquid-substrate dewetting (when the substrate viscosity $\eta_s$ is smaller than $\eta_f / \theta_e$ with $\eta_f$ the film viscosity and $\theta_e$ the equilibrium contact angle), a regime where the dewetting speed depends on the substrate thickness can be observed when such thickness becomes comparable to the rim size. In this case, moreover, the dewetting speed decreases with time, because the dissipation on the moving rim increases with the rim growth;[17] this rim-size dependent friction is also the reason for which full-slip dewetting from solid substrates is nonlinear in time.[32-34]

Iterating the process of placing one polymer layer on top of another gives rise to multilayer films that have novel application properties, including optical filtration,[35] and improved permeability.[36-40] These features have in some cases been linked to confinement and/or interfacial effects.[39, 41, 42] One method of efficiently preparing such multilayer films is called multilayer coextrusion. This industrial process involves forcing coextruded polymer flows through a series of multiplying elements, leading to materials made of thousands of alternating layers. Under typical processing conditions it is possible that all the layers have nanometric thicknesses while the total material thickness is millimetric and easily handled compared to single nanometric layers.

While multilayer coextrusion is a simple and efficient method of preparing these desired multilayer films, it has been observed for many polymer pairs that the layers break up spontaneously when reduced to a few tens of nanometers.[43-46] Such breakups, perhaps due to disjoining forces that become dominant at these length scales,[47] are often detrimental to the improved macroscopic or material properties of the obtained material.[48]

In order to study the physical phenomena responsible for such breakups and their subsequent dynamics, we recently developed a model experiment in which we observe the rupture of a polymer thin film within two thicker layers of another polymer, *i.e.* a symmetric liquid system with two polymer-polymer interfaces, instead of one polymer-polymer interface and one polymer-air interface in the case of a polymer bilayer on a rigid substrate.[49] This experiment consists of placing a thin (~100 nm) film within two thicker (≳ 1 μm) layers spin-coated on glass slides. We studied the hole formation and growth over time (*i.e.* $R = f(t)$ where $R$ is the radius of the hole and $t$ is the time) under a microscope while heating at a temperature similar to the extrusion temperature, well above $T_\mathrm{g}$. In this experiment, no external flow was applied. Though one recent article dealt with the onset of the instability in such "sandwiches" and the initial stages of hole formation (spinodal or nucleation),[50] the dynamics of such dewetting was not studied systematically. The closest experiments were those conducted by Reyssat and Quéré on the bursting of a water film within a much more viscous oil bath, even though in these experiments the viscosities involved were typically $10^4$ times lower than those involved in the polymer trilayers.[51] In such systems, a rim at the edge of the hole in the dewetting film can be observed contrary to the case of free-standing viscous films which show rimless holes and an exponential growth in time.[52-55] The resulting dynamics is then well described by making an analogy with Stokes' drag on a sphere in a viscous medium, but with the spherical geometry replaced by a cylindrical one; the dewetting rim replacing the classical falling sphere.[56, 57]

In our previous study,[49] we showed that when the matrix and film viscosities were comparable, the simple model used by Reyssat and Quéré[51] was able to capture the experimental data for a relatively large range of viscosities: the growth rate of holes, $\dot{R}$, could be estimated properly by balancing capillary and viscous forces with dissipation occurring not in the film but in the surrounding matrix (*i.e.* $\dot{R} \sim \frac{\gamma}{\eta}$ where $\gamma$ is the interfacial tension between the liquids, $\eta$ is the outer-layer viscosity, and a dot represents differentiation with respect to time). In the present article, a more systematic study takes into account the finite size of the outer layers in the trilayer system. We show that the rigid glass walls, supporting no slip, lead to a thickness-dependent dewetting rate. Simple scaling models capture well the measured hole growth dynamics, leading to a master curve which captures the dewetting data for all of the films studied.

**Materials and Methods**

A majority of the experiments presented here were conducted using commercial extrusion grades of polystyrene (PS) and polymethylmethacrylate (PMMA) that are also used for our multilayer coextrusion studies.[47, 58] Specifically, we used PS 1340 from Total, as well as PMMA VM100 and PMMA 825T supplied by Arkema (Altuglas®). To confirm results obtained using PS 1340, a non-formulated and less disperse analog has been used (weight-averaged molecular weight 244 kg/mol from Polymer Source, Inc., termed PS 244k). For commercial grades, the molecular weight distributions were characterized previously,[49] while for the PS from Polymer Source we used values provided by the manufacturer. The material properties of the main polymers used here are summarized in Table 1. Complex viscosities for PS 1340, PMMA VM100 and 825T at 180 and 200°C were previously measured.[49] As the quantities of PS 244k were too small to perform such rheological measurements, viscosities at 200 °C have been estimated using the empirical Fox-Flory equations.[59, 60]

|  | $M_w$ (kg/mol) | $M_n$ (kg/mol) | viscosity (Pa.s) |
|---|---|---|---|
| PS 1340 | 245 | 112 | 11300<br>57000 (180 °C) |
| PS 244k | 244 | 195 | 23500 |
| PMMA VM100 | 139 | 67 | 9800<br>72000 (180 °C) |
| PMMA 825T | 140 | 75 | 56900 |

Table 1. Weight-averaged, $M_w$, number-averaged, $M_n$, molecular weights and zero-shear viscosities of the polymers used in this study; unless otherwise noted, temperatures are 200 °C.

PMMA layers of thickness $H$ were prepared by spin-coating a PMMA solution onto a glass slide (Spin 150 v-3 from SPS, 25 wt% in toluene). The speed, acceleration and total rotation time were fixed at 1000 rpm, 1000 r/s$^2$ and 60 s respectively. The samples were then heated up to 160 °C for 24 hours under vacuum to remove residual stress and solvent. The thickness of the PMMA layer with these spin-coating conditions is $7.6 \pm 0.2$ µ$m$ as measured using a Veeco Profilometer (Dektak 150). Using atomic force microscopy in tapping mode (AFM, Veeco Nanoscope V; Tap300-G tips with force constant 40 N/m, resonance frequency: 300 kHz, tip radius < 10 nm from Budget Sensors, Bulgaria) we find subnanometric roughnesses on 10 µm × 10 µm scan areas. Thicknesses $0.4 < H < 12$ µm have been obtained by varying the solution concentration and spinning speed.

Thin films of PS with thickness $e$ were also prepared by spin-coating PS solutions in toluene on a silicon wafer (100 crystal orientation, from Sil'tronix). A piranha treatment was performed on the silicon wafers prior to deposition to remove organic contaminants and obtain a hydrophilic surface in order to facilitate the floating of the film onto water. Varying the concentration of PS in toluene (from 0.57 to 5.7 wt%) while keeping the speed, acceleration and rotation time of spin-coating constant (2000 rpm, 2000 r/s$^2$ and 60 s), PS film thicknesses, $e$, from 23 nm to 420 nm with subnanometric roughness have been obtained as measured using AFM. In order to check whether sample preparation affects the dewetting, several PS films were pre-annealed under vacuum at 150 °C for 24 h after spin-coating on freshly cleaved Mica substrates (V2 hi grade, 25 mm × 25 mm, from Eloise). After spin-coating, the films are cut into small pieces using a razor blade and then floated on a distilled-water bath. One of the floated PS film sections is then picked up on the PMMA substrate and left to dry for a few hours under ambient conditions.

To complete the trilayer, a second PMMA film on glass is placed on top of the bilayer at 150 °C for 3 to 5 minutes with a small force applied on top to ensure adhesion between the PMMA and PS without inducing significant flow of the polymers which could result in a change of thicknesses. The resulting samples are shown schematically in Figure 1a).

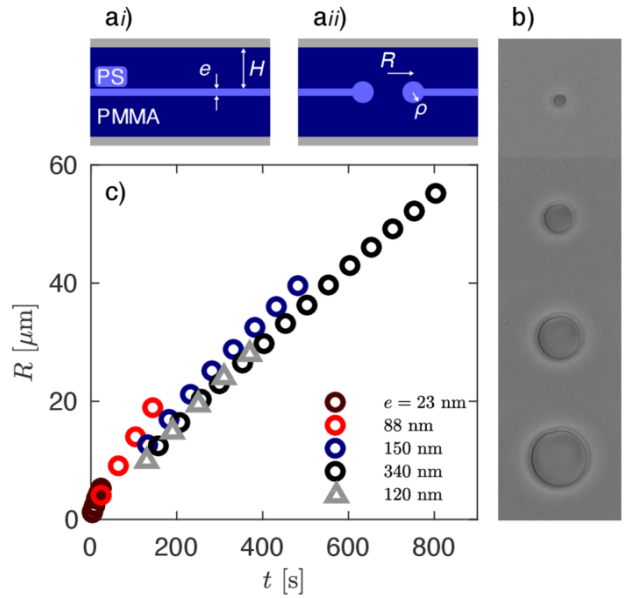

*Figure 1: a) Schematic of the experiment with lateral views i) before and ii) after hole formation and growth. b) Growth of a typical hole in a PS 1340 film within PMMA VM100 at 200 °C with e = 340 nm, H = 7.6 µm and times 610 ≤ t ≤ 1210 s from top to bottom; each frame is 230 µm wide. c) Hole radius as a function of time for different thicknesses, e, of PS 1340 in PMMA VM100 at 200 °C for H = 7.6 µm (circles) or H = 3 µm (triangles).*

These trilayer systems were then placed in a Mettler FP80 heating stage already set to the chosen temperature under an Olympus BH-2 optical microscope with a 20× magnification (or 5× for the thicker films where larger holes were studied). Images were recorded at regular intervals depending on the length of the experiment (minutes to hours) to observe the appearance of holes in the PS and their growth over time.

For each studied hole, at least 4 images as shown in Figure 1b) have been analyzed, the inner diameter of the holes being measured using Olympus analysis software with a typical precision of ± 0.5 µm. For each dewetting film in a trilayer, furthermore, the growth of at least three independent holes is studied. Each hole is chosen so that there is no surrounding hole closer than typically the diameter of the studied hole; the possible interactions between holes are therefore neglected. At least two independent dewetting films in trilayers are studied for each experimental condition (*e.g.* film thickness combination and temperature) such that each quoted value of $\dot{R}$ in the following represents at least six independent measurements of the hole dynamics.

**Results**

As shown in Figure 1b), a circular hole of radius $R$ grows with time $t$ following a nucleation event. The dewetted PS is collected into a rim, which is assumed to be circular in cross section with a radius $\rho$, see part *ii*) of Figure 1a) and Figure 1b). In Figure 1c) we show the temporal evolution of the hole radius for several trilayer hole growth experiments, suggesting in general that the hole growth rate depends on film thickness, $e$, and outer layer thickness, $H$.

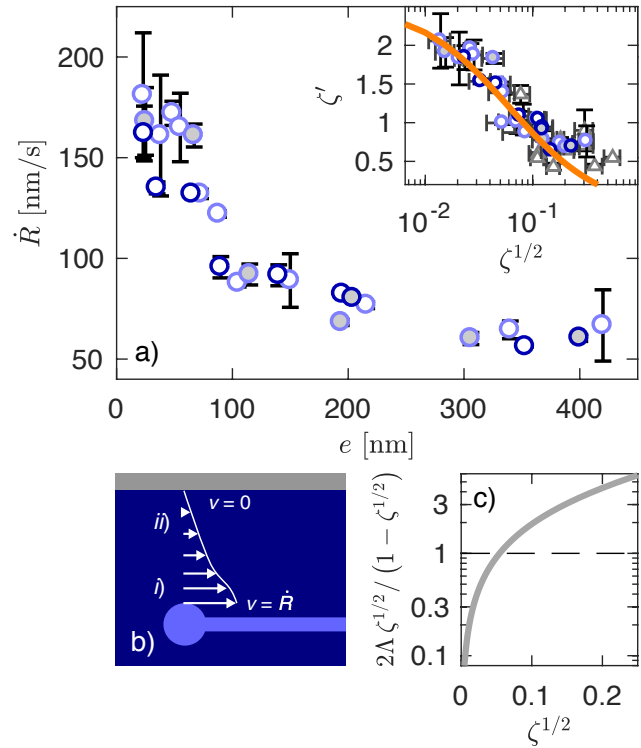

*Figure 2: a) Hole growth rate, $\dot{R}$, of PS in PMMA VM100 as a function of PS film thickness, $e$, for PMMA films with constant thickness $H = 7.6 \pm 0.2$ μm and at temperature $T = 200$ °C. Grey-filled symbols denote those for which the PS films were pre-annealed before transfer to the PMMA outer layer; the color code is PS 1340 (light blue), PS 244k (blue). The inset shows the dimensionless growth rate as a function of the square root of the dimensionless hole radius (Equation (7)) for each experimental condition studied including data using PMMA 825T at 200 °C, and PMMA VM100 at 180 °C, with thicknesses $0.4 \leq H \leq 12$ μm and $50 \leq e \leq 280$ nm. The orange line is the best fit for the data restricted to $\zeta^{1/2} \leq 0.18$ using Equation (6). b) Schematic of the scaling regions with i) the Stokes-like flow associated with a cylinder and ii) the Couette-like flow associated with the motion of the rim near the wall. c) Evaluation of the Couette contribution to the dissipation in Equation (6) as a function of square root of the dimensionless hole radius.*

First, we investigate the growth rate as a function of PS film thickness, $e$, as shown in Figure 2a) for different PS, annealed or not, with $H = 7.6$ µm and using PMMA VM100 at $T = 200$ °C. As a rate we take the slope of the best linear fit to all of the data for a given series of $R(t)$.[61] In this main panel of Figure 2a), we hold $H, T$ and the molecular weight of the PMMA constant, while $e$ is varied systematically for two different PS interiors.

The data in Figure 2a) show that $\dot{R}$ increases with decreasing film thickness, but that it does not depend strongly on the PS viscosity or the processing conditions (*i.e.* pre-annealed PS or not). We note that the zero-shear viscosity ratio at constant PMMA viscosity, $\frac{\eta}{\eta_{0\,PS}}$, varies between 0.4 and 0.9 (Table 1), which is in the range of previously studied trilayers for which we found that only the PMMA viscosity was controlling the dynamics.[49] Complementary experiments with 64 and 835 kg/mol PS in PMMA VM100 confirm the weak dependence of the growth rate on the PS film viscosity.[*] Henceforth we thus consider only the viscosity of the outer PMMA matrix, $\eta$.

Based on the fact that the hole growth rate increases as the film becomes thinner, one could make the hypothesis that our experiments reveal a confinement effect within the PS. Indeed, the PS thickness at which the rate begins to strongly increase in Figure 2a) is roughly 50 nm. This value: is only slightly above the thickness at which apparent glass transition temperature reductions begin to be observed;[4, 6, 62] is a typical value of slip lengths for PS melts on certain hydrophobic coatings;[63-65] and approaches the length scales for which disjoining forces may be operative.[66-68] However, invoking either slip or disjoining forces does not allow to quantitatively rationalize the data. In contrast, we show in the following that the observed increase in the dewetting rate is well explained by the presence of the glass slides which engender a no-slip boundary condition, despite being several microns away from the dewetting film of tens to hundreds of nanometers thickness.

---

[*] Using estimations from the Fox-Flory equations[59, 60] the PS viscosity spans nearly 4 orders of magnitude, from 250 Pa.s for PS 64k to $1.6 \times 10^6$ Pa.s for 835K at 200 °C. These polymers were also purchased from Polymer Source, having polydispersities 1.04 and 1.16. While the capillary driven dynamics is often inversely proportional to the viscosity,[1, 28] the dewetting of the PS 835k for $e = 350$ nm is only 5 times slower than the PS 244k for nearly two orders of magnitude change in viscosity; similarly the PS 64k hole growth rate is only roughly 30% higher than the PS 1340 and PS 244k for a similar change in the viscosity and at the same thickness.

**Model**

In order to model the growth of holes in trilayer films, we use a scaling argument involving the viscous power dissipation balanced by the capillary driving power. We assume that dewetted material is collected into a toroidal rim with a circular cross section of radius $\rho$ and a toroidal radius $R$, which is a reasonable approximation in the early stages for which $\rho \ll H$. Volume conservation then implies a connection between $\rho$ and $R$ through $\pi R^2 e = 2\pi R \pi \rho^2$. On undergoing an infinitesimal increase, $dR$, in the radius, the capillary energy, $U$, decreases by an amount

$$dU = \gamma dA, \tag{1}$$

$$\sim 4\gamma R \, dR, \tag{2}$$

where $\gamma$ is the PS-PMMA surface tension, $A = 2 \cdot \pi R^2$ for the two PS-PMMA surfaces that disappeared in the process of creating a cylindrical dewetting hole of radius $R$. In Equation (2) we neglect the excess area around the rim which is valid in the limit for which $R \gg e$, as is always satisfied in the experiments, and we retain the numerical factor 2 representing the two surfaces.

The case of $H \rightarrow \infty$ was treated by Reyssat and Quéré,[51] and can be recovered by considering the motion of a rigid cylinder in a viscous matrix. The viscously dissipated power generally reads[69] $\mathcal{P} = \frac{\eta}{2} \int d\Omega \left( \partial_{x_i} v_j + \partial_{x_j} v_i \right)^2$ where $\Omega$ is the fluid volume, $v_j$ is the fluid velocity in the $j$-direction and $\partial_{x_i}$ denotes differentiation with respect to the coordinate in the $i$-direction, $x_i$. Assuming a Stokes-like fluid flow around a rigid cylinder, the fluid velocity $\dot{R}$ decays over a distance $\rho$, see region *i*) of Figure 2b). Recalling the toroidal geometry, one gets the following scaling for the dissipated power:

$$\mathcal{P}_{\text{Stokes}} \sim \eta \left( \frac{\dot{R}}{\rho} \right)^2 \rho^2 R, \tag{3}$$

where the term in brackets is the typical velocity gradient and that to the right is the typical volume. Setting $\frac{dU}{dt} = \mathcal{P}_{\text{Stokes}}$, meaning that all the energy gained by capillarity is viscously dissipated, we recover a constant growth rate dewetting, $\dot{R} \sim \gamma/\eta$. While we have here neglected the weak logarithmic term evoked by Reyssat and Quéré applying to a cylinder in an infinite bath,[51, 56, 57] in the finite case[70, 71] this term does not appear in the analytic expressions for the Stokes drag on a cylinder. We expand on this statement in the appendix, where a comparison is made between our scaling arguments and those incorporating analytic approximations for the drag on a cylinder in a slit.

Now considering the glass walls bounding the PMMA layers, we note that the fluid velocity goes to zero there under the assumption of a no-slip boundary condition. Since the fluid velocity at the top of the dewetted rim is of order $\dot{R}$, there is a second typical velocity gradient and thus a Couette-like dissipation associated with the motion of the rim, see region *ii*) of Figure 2b). In that region, the dissipated power scales as

$$\mathcal{P}_{\text{Couette}} \sim 2\eta \left(\frac{\dot{R}}{H-\rho}\right)^2 \rho R (H-\rho), \tag{4}$$

where the factor of two accounts for the fact that there are two PMMA layers. Adding up the two dissipation powers, Equations (3) and (4), and balancing this sum with the capillary driving power, $dU/dt$, gives the growth rate as

$$\dot{R} = \omega \frac{\gamma}{\eta} \left[1 + 2\Lambda \frac{\rho}{H-\rho}\right]^{-1}, \tag{5}$$

where $\omega$ and $\Lambda$ are unknown numerical prefactors due to the scaling approach. Using the volume conservation constraint above, Equation (5) can be rewritten as

$$\frac{\eta \dot{R}}{\gamma} = \zeta' = \omega \left[1 + 2\Lambda \frac{\zeta^{1/2}}{1-\zeta^{1/2}}\right]^{-1}, \tag{6}$$

where we have introduced the dimensionless hole radius

$$\zeta = \frac{eR}{2\pi H^2}, \tag{7}$$

and the dimensionless time

$$\tau = \frac{\gamma e t}{2\pi \eta H^2}, \quad (8)$$

the derivative with respect to which is indicated by a prime. We expect this expression to be valid as long as the rim size does not become too close to the PMMA film thickness. Indeed, the model would predict that no dewetting can be observed once the rim touches the wall, which is not generally the case since a viscous wetting film[68, 72, 73] may be formed and the rim shape would deviate significantly from the idealized circular cross section we have assumed.

**Discussion**

As a test of the model we show in the inset of Figure 2a) the measured growth rate, $\zeta'$, as a function of the square root of the dimensionless hole radius, $\zeta^{1/2}$; equivalently, this is the dimensionless cylinder radius. For the purposes of this analysis, we have replaced $R$ in Equation (7) with its average over the limited experimental range for a given hole growth experiment. The vertical and horizontal error bars represent the spread of values for a given combination of $\{e, H, T\}$ and molecular weights. In addition to the data taken from the main part of the figure, the inset of Figure 2a) contains data using PMMA 825T at 200 °C, and PMMA VM100 at 180 °C (thus changing $\eta$ as shown in Table 1), and a range of PMMA film thickness $0.4 \leq H \leq 12$ µm is shown; these data are all represented using grey triangles. For the experimental conditions involving different PS and PMMA film thicknesses, different processing conditions and temperatures, all the data fall on the same curve. In addition, by fitting to Equation (6), the two free parameters are determined as $\omega = 2.4 \pm 0.3$ and $\Lambda = 8 \pm 2$; that these values are of order unity suggests that the scaling approach is simply missing geometrical prefactors. Note that the fitting was restricted to small values of $\zeta^{1/2} \leq 0.18$ allowing to capture well all the data at small $\zeta$. As a complementary check on its relative importance, in Figure 2c) we show the term associated to the Couette-like flow in Equation (6). There we see that for $\zeta^{1/2} \lesssim 0.05$, *i.e.* when the rim is far from the wall, $\mathcal{P}_{\text{Couette}}$ is negligible with respect to $\mathcal{P}_{\text{Stokes}}$. However, there is a crossover and

$\mathcal{P}_{\text{Couette}}$ becomes larger than $\mathcal{P}_{\text{Stokes}}$ when $\zeta^{1/2} > 0.05$; when the rim size becomes larger than just 5% of the channel width the Couette-like dissipation becomes dominant. This observation provides a guide as to when liquid baths may be considered infinite.

To further test the model, we note that each data point in Figure 1c) and those used to construct Figure 2 are expected to follow a master curve if the hole radius and time are non-dimensionalized according to Equations (7) and (8); all the associated parameters have been measured in independent experiments. Equation (6) was integrated using the MATLAB ode45 routine with the parameters $\omega = 2.5$ and $\Lambda = 8$ obtained from the fit shown in the inset of Figure 2a), using $\zeta(0) = 0$ as an initial condition. In Figure 3, we show the numerical solution of Equation (6) along with the experimental points $\zeta(\tau)$. Experimentally, we included a small (compared to the experimental duration) offset time to ensure that each experimental curve extrapolates to $R = 0$ at $t = 0$. Following this non-dimensionalization and offsetting procedure, all of the experimental data fall onto the same master curve. The data comprises trilayers with PS and PMMA film thicknesses in the ranges $23 < e < 420$ nm and $2.6 < H < 12$ μm, and capillary velocities ranging from $15 < \frac{\gamma}{\eta} < 90$ nm/s. Moreover, the collapsed experimental results conform to the scaling prediction over nearly four orders of magnitude in dimensionless time and radius. This agreement suggests that the model incorporates the appropriate physical mechanisms for the observed thickness dependence of the hole growth rate seen, in particular, in Figure 2a). We note finally that deviations from the scaling prediction occur only for the thickest PS films and thinnest PMMA films studied. These correspond to a regime for which $\rho \to H$, the rim size becomes comparable to the PMMA film thickness. As discussed above, we expect that in this regime the idealized circular cross section of the rim is an invalid hypothesis. The asymptotic regime for which $\zeta \gg 1$, i.e. for which $R \gg H^2 e^{-1}$, could correspond to either: *i*) wetting of the glass by PS and a corresponding contact line motion;[28] or *ii*) a thin lubricating layer reminiscent of the films of Bretherton[72] and others.[68, 73]

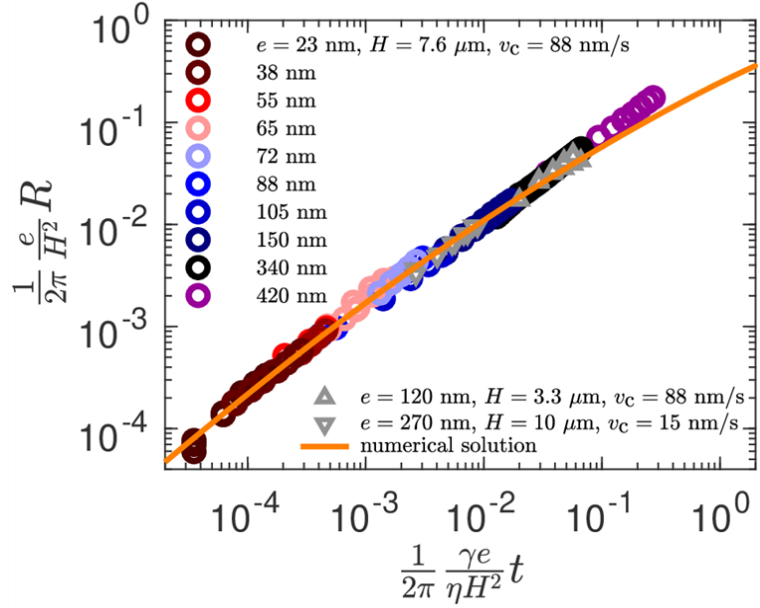

*Figure 3: Dimensionless hole radius as a function of dimensionless time for PS films, with indicated thickness e in nm, dewetting from symmetric PMMA layers for all of the experimental conditions studied. The solid orange line represents the numerical solution of Equation (6), using the fit parameters obtained in Figure 2: $\omega = 2.5$ and $\Lambda = 7$.*

**Conclusion**

In this study we have demonstrated that the growth rate of holes in a PS thin film sandwiched between two PMMA outer layers can be strongly influenced by the chosen outer layer thickness, even while the latter thickness is orders of magnitude larger than that of the PS layers. In fact, to predict accurately the growth rate one must take into account the distance between the dewetting rim and the glass slides on which the outer layers are deposited. In our scaling model, we balance capillary driving with viscous dissipation; the latter, considered within the outer layers only, includes contributions from both classical Stokes-like and Couette-like flow with a no-slip condition at the glass boundary. This approach captures well the hole growth rate in thin films over a large range of experimental conditions, including changes in viscosity, as well as in the thicknesses of the thin film and of the outer layers. This experimental system may be used as a rheological tool to study viscosities in nanolayers, or to measure the interfacial tension between polymer melts. Furthermore, by extending the trilayer technique to systems with specific boundaries of interest (*i.e.* soft or slippery), we expect that the dewetting rim will serve as a symmetric, contactless surface probe. Finally, this model configuration might provide practical insights for the puzzling nanoscale phenomena at play in industrial nanolayer coextrusion processes.

**Acknowledgements**

The authors warmly thank F. Restagno and K. Kadri for useful discussions, Y. Zhu for performing several of the dewetting experiments presented in this manuscript, and gratefully acknowledge Arkema for providing the PMMA VM100 and 825T. J.D.M. was supported by LabEX ENS-ICFP: No. ANR-10- LABX-0010/ANR-10- IDEX-0001-02 PSL. T.S. thanks the Global Station for Soft Matter, a project of Global Institution for Collaborative Research and Education at Hokkaido University.

**Appendix**

In this appendix we compare the scaling model of Equations (1) – (6) to similar arguments incorporating analytical predictions of the Stokes drag on a straight, infinite cylinder in a viscous liquid centered between two parallel plates. For notational consistency, we choose the cylinder to have radius $\rho$, and the slit width to be $2H$. We first present the case for a cylinder being dragged by a constant force in an otherwise quiescent fluid, and then the case in which a cylinder is placed in a pressure driven (*i.e.* Poiseuille) flow.

Takaisi[70] considered the case of an infinite cylinder with constant force applied. The resulting relation gives the drag force per unit length of the cylinder, $D_C$, as a linear function of the velocity

$$D_C = \frac{4\pi\eta U}{\ln\left(\frac{H}{\rho}\right) - \Lambda_T}, \qquad (A1)$$

where $U$ is the cylinder velocity parallel to the slit walls and Takaisi predicted $\Lambda_T \approx 0.9156$. Making the substitution of the volume conservation constraint, $Re = 2\pi\rho^2$, identifying the cylinder velocity with the dewetting rim speed, $U = \dot{R}$, and balancing the viscous drag with the driving force per unit length (that is, the surface tension), $2\gamma$, we obtain the relation

$$\zeta'_C = \omega_C \left( \ln\left(\zeta^{-\frac{1}{2}}\right) - \Lambda_C \right), \tag{A2}$$

where $\omega_C$ is an unknown prefactor due to the scaling approach, and we have used the dimensionless variables defined in Equations (7) and (8).

A cylinder instead driven by a pressure gradient flow exhibits a different force/velocity relation. Richou *et al.*[71] studied this case and approximate the drag force per unit length of the cylinder as

$$D_P = \frac{4\pi\eta U\left(\left(\frac{\rho}{H}\right)^2 - 2\right)}{P_0 - P_1\left(\frac{\rho}{H}\right)^2 + 2\ln\left(\frac{\rho}{H}\right)}, \tag{A3}$$

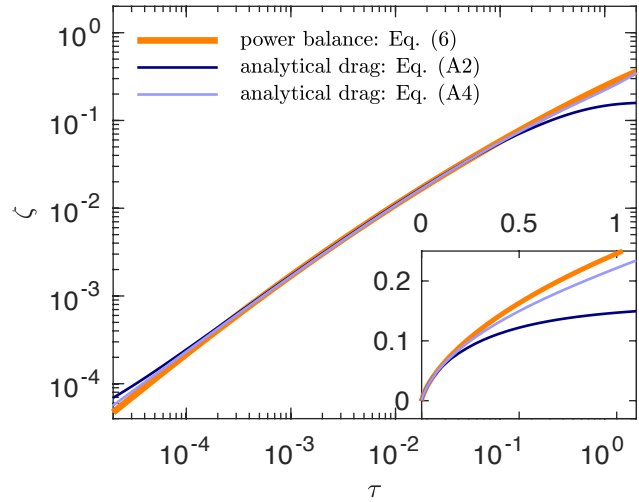

*Figure A1: Predictions of the scaling models for the dimensionless dewetting dynamics. For the scaling model from a power balance (orange) we have used the same fitting parameters discussed in the main text. The prefactors for models leading to Equations (A2) and (A4) (dark and light blue) are identical with $\omega_C = \omega_P = 0.65$. The inset represents the same results on linear axes.*

valid for $\rho/H < 0.4$, and where $P_0 \approx 1.9362, P_1 \approx 3.7520$. Making the same substitutions as for the constant-force case above, we find the equation of motion

$$\zeta'_P = \omega_P \frac{P_0 - P_1\zeta + 2\ln(\zeta^{1/2})}{(\zeta - 2)}, \quad (A4)$$

where $\omega_P$ is also a geometrical prefactor.

In Figure A1 we show the scaling result predicted by Equation (6) as well as the results incorporating analytical drag approximations in Equations (A2) and (A4). These latter predictions were obtained by numerical integration again using the MATLAB ode45 routine. In Figure A1 it is shown that over the range of experimentally observed $(\zeta, \tau)$ accessed in Figure 3, neither of the predictions distinguishes the data better than the others, although the numerical integration of Equation (A2) deviates from the data more strongly than the predictions of Equations (6) and (A4). The similarity of the scaling models incorporating analytical estimations of the cylinder drag and our simple power balance suggests further that the latter (i.e. Equations (1)-(6)) capture the essential mechanisms operative in the dewetting experiments. We furthermore note that the fitting parameters in both cases of Equations (A2) and (A4) are the same and we find $\omega_C = \omega_P = 0.65$ suggesting indeed that missing prefactors are purely geometric.

* Addresses to which correspondence should be addressed: joshua.mcgraw@espci.fr, guillaume.miquelardgarnier@lecnam.net